\documentclass[final,3p,times]{elsarticle}

\usepackage{lineno,hyperref}

\usepackage{upgreek}
\usepackage[parfill]{parskip}
\usepackage[margin=5pt,font=small,labelfont=bf]{caption}
\usepackage{natbib}
\usepackage[none]{hyphenat}
\usepackage{amsthm}
\usepackage{amsmath}
\usepackage{array}
\usepackage{color}
\usepackage{multirow}
\usepackage{graphics}
\usepackage{wrapfig}
\usepackage{float}
\usepackage{placeins}
\usepackage{rotating,booktabs}
\usepackage{multirow}
\usepackage{lipsum}

\journal{Icarus}

\biboptions{authoryear}

\begin{document}

\begin{frontmatter}


\title{Long-term variability of Jupiter's northern auroral 8-$\upmu$m CH$_4$ emissions}
\author[jpl]{J. A. Sinclair}\corref{cor1}
\ead{james.sinclair@jpl.nasa.gov}
\author[jpl]{R. West}
\author[niss,integra]{J. M. Barbara}
\author[nict]{C. Tao}
\author[jpl]{G. S. Orton}
\author[swri]{T. K. Greathouse}
\author[swri]{R. S. Giles}
\author[liege]{D. Grodent}
\author[leic]{L. N. Fletcher}
\author[oxf]{P. G. J. Irwin}

\address[jpl]{Jet Propulsion Laboratory/California Institute of Technology, 48O0 Oak Grove Dr, Pasadena, CA 91109, United States}
\address[niss]{NASA Goddard Institute for Space Studies, 2880 Broadway, New York, NY 10025, United States}
\address[integra]{Autonomic Integra LLC, 2880 Broadway, New York, NY 10025, United States}
\address[nict]{National Institute of Information and Communications Technology, 4-2-1, Nukui-Kitamachi, Koganei, Tokyo 184-8795, Japan}
\address[swri]{Southwest Research Institute, 6220 Culebra Rd, San Antonio, TX 78238, United States}
\address[liege]{Université de Liège, STAR Institute, Laboratoire de Physique Atmosphérique et Planétaire, Quartier Agora allée du six Août 19 c 4000 Liège 1, Belgium}
\address[leic]{School of Physics \& Astronomy, University of Leicester, University Road, Leicester, LE1 7RH, United Kingdom}
\address[oxf]{Atmospheric, Oceanic \& Planetary Physics, University of Oxford, Parks Road, Oxford, OX1 3PU, United Kingdom}

\begin{abstract}
We present a study of the long term variability of Jupiter's mid-infrared CH$_4$ auroral emissions.  7.7-7.9 $\upmu$m images of Jupiter recorded  by NASA's Infrared Telescope Facility, Subaru and Gemini-South over the last three decades were collated in order to quantify the magnitude and timescales over which the northern auroral hotspot's CH$_4$ emission varies.  These emissions predominantly sound the 10- to 1-mbar pressure range and therefore highlight the temporal variability of lower-stratospheric auroral-related heating.  We find that the ratio of the radiance of the poleward northern auroral emissions to a lower-latitude zonal-mean, henceforth `Relative Poleward Radiance' or RPR, exhibits varability over a 37\% range and over a range of apparent timescales.  We searched for patterns of variability in order to test whether seasonally varying solar insolation, the 11-year solar cycle, or short-term solar wind variability at Jupiter's magnetopause could explain the observed evolution.  The variability of the RPR exhibits a weak (r $<$ 0.2) correlation with both the instantaneous and phase-lagged solar insolation received at Jupiter's high-northern latitudes.  This rules out the hypothesis suggested in previous work (e.g. \citealt{sinclair_2017a,sinclair_2018a}) that shortwave solar heating of aurorally produced haze particles is the dominant auroral-related heating mechanism in the lower stratosphere.  We also find the variability exhibits negligible (r $<$ 0.18) correlation with both the instantaneous and phase-lagged monthly-mean sunspot number, which therefore rules out a long-term variability associated with the solar cycle.   On shorter timescales, we find moderate correlations of the RPR with solar wind conditions at Jupiter in the preceding days before images were recorded.  For example, we find correlations of r = 0.45 and r = 0.51 of the RPR with the mean and standard deviation solar wind dynamical pressure in the preceding 7 days.  The moderate correlation suggests that either: 1) only a subset of solar wind compressions lead to brighter, poleward CH$_4$ emissions and/or 2) a subset of CH$_4$ emission brightening events are driven by internal magnetospheric processes (e.g. Io activity) and independent of solar wind enhancements.

\end{abstract}


\end{frontmatter}

\section{Introduction}

Auroral processes on Jupiter represent an extreme example of space weather due to the planet's large magnetic field and the internal plasma source in the form of Io's volcanism.  Solar wind perturbations to the magnetosphere and internal magnetospheric dynamics accelerate energetic ions and electrons into Jupiter's polar atmosphere producing auroral emissions over a large range of the electromagnetic spectrum (e.g. \citealt{dunn_2017}, \citealt{hue_2021}, \citealt{odonoghue_2021}).  A significant amount of energy is  ultimately deposited as deep as Jupiter's lower stratosphere ($\sim$1 mbar), heating the atmosphere (e.g. \citealt{drossart_1993,sinclair_2018a}), enriching the abundances of higher-order hydrocarbons (e.g. \citealt{sinclair_2017a,sinclair_2019a}), thereby producing enhanced mid-infrared emission features of stratospheric CH$_4$ and its photochemical by-products species (e.g. \citealt{caldwell_1980,kim_1985, kostiuk_1993,flasar_2004_jet}).  

\begin{figure*}[t!]
\begin{center}
\includegraphics[width=0.8\textwidth]{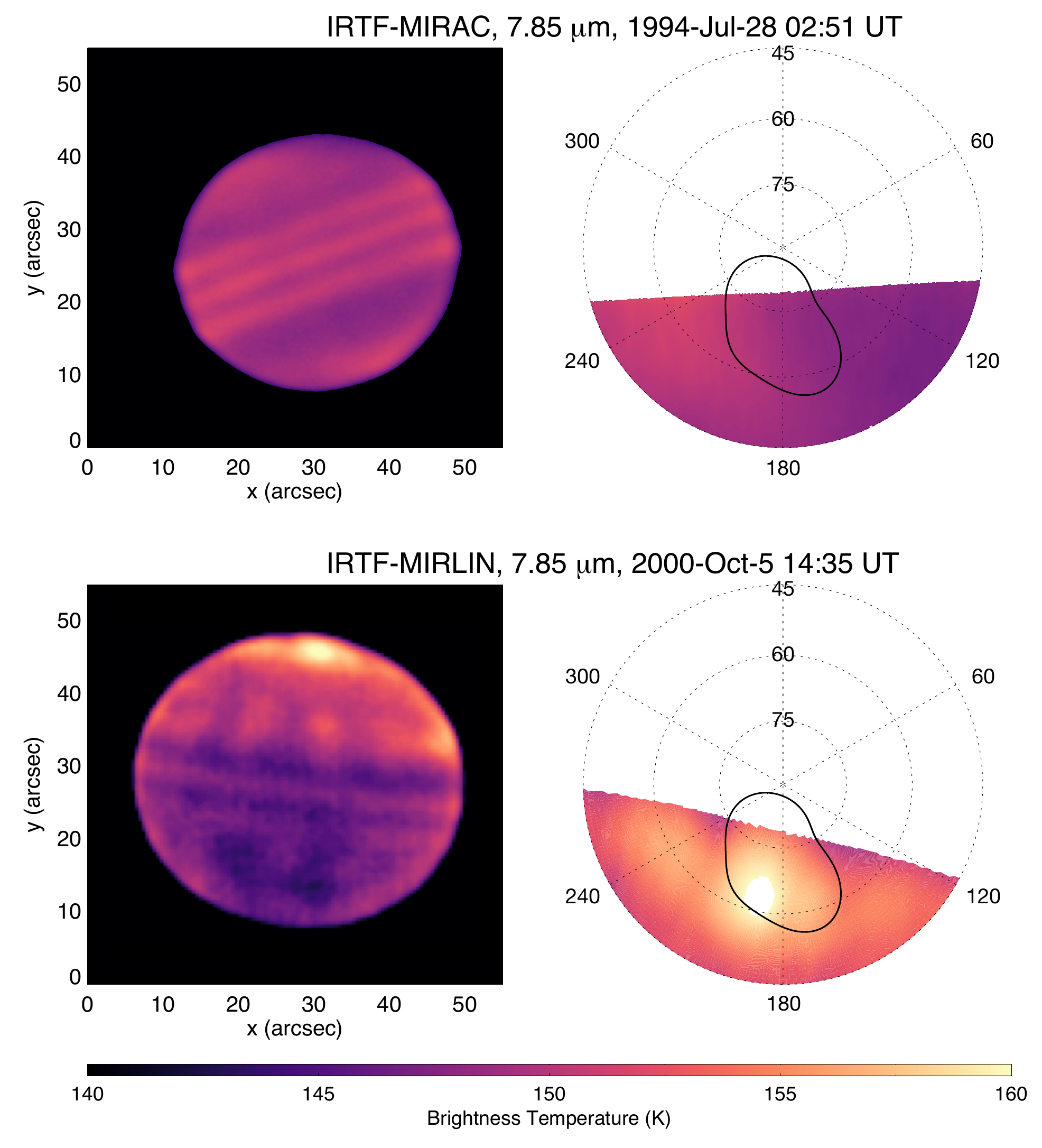}
\caption{The left column shows example images of of Jupiter recorded using the circular-variable filter (CVF) centered at 7.85 $\upmu$m.  The right column shows the corresponding polar projections with the solid, black line denoting the statistical-mean position of the ultraviolet main auroral emission \citep{bonfond_2017}.  All images are in brightness temperature units according the colorbar.   }
\label{fig:example_images}
\end{center}
\end{figure*}

The auroral hotspots in both auroral regions have been observed using both Earth-based (e.g. \citealt{kostiuk_1993,drossart_1993,livengood_1993,kim_2014,sinclair_2018a}) and spacecraft instrumentation \citep{flasar_2004_jet,sinclair_2017a,sinclair_2019b}.  Generally, the strongest mid-infrared auroral emissions, commonly termed the `auroral hotspot', are observed spatially-coincident with the `poleward' ultraviolet auroral emissions: enclosed within or magnetospherically-poleward of the main auroral oval (e.g. \citealt{grodent_2015}).   Anecdotally, the auroral hotspots have been observed to vary in brightness, disappear and re-emerge but the mechanisms driving this variability have remained uncertain.  This is demonstrated in Figure \ref{fig:example_images}, which shows examples where the northern auroral hotspot is absent and another where it is present.  

In previous studies, inversions of the CH$_4$ emission spectra in order to derive the vertical temperature profile have revealed that the enhanced mid-infrared emissions result from, in part, two, vertically-discrete layers of heating in the lower ($\sim$1 mbar) and upper ($<$ 0.1 mbar) stratosphere \citep{kostiuk_agu_2016,sinclair_2017a,sinclair_2017b,sinclair_2018a,sinclair_2020b}.  The upper-stratospheric heating is vertically coincident with the ultraviolet auroral emissions and is therefore expected to arise from the same processes that produce the ultraviolet auroral emissions: chemical heating, H$_2$ dissociation from excited states \citep{grodent_2001}, joule heating from Pedersen currents (e.g. \citealt{badman_2015}) and ion drag.  These processes essentially move the base of the thermosphere to lower altitudes, as has been shown using general circulation modelling of Jupiter's thermosphere  \citep{bougher_2005}.

The mechanism(s) responsible for the deeper 1-mbar heating has/have remained uncertain though several hypotheses have been suggested in previous work.  One suggestion was the 1-mbar heating results from shortwave solar heating of haze particles \citep{sinclair_2017a,sinclair_2018a} produced by complex chemistry prevalent in Jupiter's auroral regions (e.g. \citealt{friedson_2002},\citealt{wong_2003}).   However, the results of this study rule out this hypothesis as the dominant source of heating, as detailed further in the Discussion.  

Direct precipitation and heating of a higher-energy population of charged particles from the magnetosphere has also been suggested as the source of the lower-stratospheic heating (e.g. \citealt{sinclair_2017a,kim_2017,kim_2023}).  However, at the time of writing, we consider this mechanism less viable for the following reasons.  Firstly, ion and electron precipitation modelling demonstrates that negligible energy directly reaches pressures higher than $\sim$0.1 mbar (e.g. \citealt{ozak_2010,gustin_2016,houston_2018}).  Electrons would have to be accelerated to near-relativistic, $>$1 MeV energies in order to precipitate at 1 mbar and \textit{in situ} Juno measurements have not detected a sufficient downward component at such energies (e.g. \citealt{clark_2017,paranicas_2018,mauk_2020}).  It is possible that an acceleration region exists below the altitudes of the Juno spacecraft in previous orbits.  Measurements by Juno during its extended mission, where the northern polar region will be crossed at increasingly lower altitudes, may be able to determine whether such an acceleration region exists \citep{greathouse_2021}.  Secondly, auroral-related heating is observed at $\sim$1 mbar and pressures lower than 0.01 mbar, with relatively less heating at the intermediate $\sim$0.1 mbar level (e.g. \citealt{kostiuk_agu_2016,sinclair_2018a,sinclair_2020b,sinclair_2023b}).  For this pattern of heating to be explained by direct precipitation of electrons, there would have to be higher fluxes of 100-200 keV and 1-2 MeV electrons, which would respectively precipitate at 0.01 mbar and 1 mbar, compared to $\sim$500 keV electrons, which would precipitate at 0.1 mbar \citep{gustin_2016}. As far as we are aware, this has not been observed in \textit{in situ} Juno measurements: if anything, the energy spectra of downward electrons peaks in the 100 - 500 keV range (e.g. \citealt{paranicas_2018}).

Most recently, \citet{cavalie_2021} observed counterrotating winds at $\sim$0.1 mbar coincident with the main auroral emissions, which they suggested were produced by ion-neutral collisions.  This is possibly an extension of a similar counterrotating electrojet predicted and observed at thermospheric altitudes (e.g. \citealt{achilleos_2001,johnson_2018}). Atmospheric subsidence associated with this circulation may produce adiabatic heating poleward of the main auroral emission and at pressures higher than 0.1 mbar and thus may be responsible for the 1-mbar auroral-related heating. The fact that a temperature minimum is observed at $\sim$0.1 mbar, as well as the fact that strong H$_3^+$ emissions are not observed poleward of the main auroral emissions, would suggest the atmospheric subsidence is limited to the lower stratosphere (p $<$ 0.1 mbar).  Dynamical modelling of this potential mechanism is required to explore whether it is the source of the 1-mbar auroral-related heating. 

In this work, our goal was to determine the magnitude and timescales over which Jupiter's CH$_4$ mid-infrared auroral hotspot varies using data over the largest time range possible.  We were motivated to perform this analysis for the following reasons.  First, we would quantify the variability that has been observed anecdotally for decades.  Second, in searching for patterns of temporal variability or lack thereof, we would be able to test the aforementioned hypotheses of the source of Jupiter's 1-mbar auroral-related heating.  For example, if shortwave solar heating of haze particles is indeed the dominant source of heating at 1 mbar, given the $\sim$3 year radiative cooling lifetimes at this altitude (e.g. \citealt{zhang_2013a}), one would expect the mid-infrared emissions from the auroral hotspot to vary in accordance with seasonally varying solar insolation over the course of a Jupiter year (12 Earth years) due to its $\sim$3$^\circ$ axial tilt.  

We chose to focus our study only on the northern auroral hotspot ($>$55$^\circ$N, 150 - 210$^\circ$W, see Figure \ref{fig:example_images}) since it is at a relatively lower latitude and therefore more observable from Earth-based observatories compared to the southern auroral hotspot. While the magnitude of CH$_4$ emissions of the northern auroral hotspot have been quantified in many previous studies over a collectively large time range (e.g. \citealt{caldwell_1980,kim_1985,flasar_2004_jet,sinclair_2018a}), differences in spatial resolution, spectral resolution, methods of reduction and analysis between different studies would hinder or complicate any interpretation of variability.  Using emissions of C$_2$H$_6$ of the auroral hotspot recorded over a large time range, \citet{kostiuk_agu_2016} observed more variability of C$_2$H$_6$ emission of Jupiter's northern auroral oval during periods of solar maxima compared to solar minima.   However, the photochemical lifetimes of CH$_4$ and C$_2$H$_6$ at a given altitude differ greatly (e.g. \citealt{moses_2005,hue_2018}) and so the temporal behavior of each molecule would also be expected to differ.  We were therefore motivated to perform a robust and consistent analysis of data capturing Jupiter's northern auroral hotspot in CH$_4$ emission in order to quantify the magnitude and timescales over which variability occurs.  To this goal, we chose to focus our analysis on Earth-based broadband imaging of Jupiter's CH$_4$ emission in filters centered between 7.7-7.9 $\upmu$m for the following reasons.  First, there is an extensive record of broadband 7.7-7.9 $\upmu$m images of Jupiter, which improves the potential temporal sampling and particularly the time range of the study, in comparison to spectroscopic observations.  Second, broadband imaging in this filter range is predominantly more sensitive to the lower stratosphere (10 $<$ p $<$ 0.1 mbar), as demonstrated in Section \ref{sec:rad}, which allows us to the test aforementioned hypotheses for the 1-mbar auroral heating.  

\section{Observations}\label{sec:obs}

We searched an extensive record of 7 - 8 $\upmu$m images of Jupiter recorded at NASA's Infrared Telescope Facility, the Subaru telescope (both at the summit of Mauna Kea), Gemini-South at Cerro Pachón, and the Very Large Telescope (VLT) at Cerro Paranal.  We omitted observations recorded at airmasses of greater than 2.2.  We also focused on images recorded at sub-observer longitudes in the 150 - 210$^\circ$W (System III) longitude range, such that the northern auroral hotspot is on or near the central meridian.  While several VLT images did satisfy these conditions, the reduction of images available at the time of the study had overlap of the target and sky frames, which obscured higher-northern latitudes, and were therefore omitted from this work.  The remaining 33 observations that satisfy the aforementioned criteria are detailed in Table \ref{tab:obs} in chronological order.

\subsection{Imaging}

At NASA's Infrared Telescope Facility (IRTF), 7.7 - 7.9 $\upmu$m images of Jupiter were recorded using the Mid-Infrared Array Camera, MIRAC \citep{hoffmann_1993} from 1994 to 1997, the Mid-Infrared Large-Well Imager, MIRLIN \citep{ressler_1994} from 1998 to 2001 and the Mid-Infrared Spectrometer and Imager, MIRSI \citep{deutsch_2003} from 2003 to 2010.  

Images of Jupiter at 7.8 $\upmu$m were acquired using the COMICS instrument \citep{kataza_2000} at the 8-meter Subaru telescope between 2008 and 2020. The COMICS detector has a pixel scale of 0.13'' and a field-of-view (FOV) of 45'' x 32''.  The FOV could not capture the entire disk of Jupiter (37 - 45") in a single exposure and so, 2 or 4 individual images were recorded offset from the center of the disk and subsequently mosaiced together during the reduction.  Individual halves or quadrants were positioned such that they overlapped, which also served to remove detector artefacts.  The Thermal-Region Camera and Spectrograph, T-ReCS \citep{debuizer_2005} is a mid-infrared imager and spectrograph mounted on the 8-meter Gemini South telescope at Cerro Pachón, Chile.  Ultimately, only one T-ReCS' image recorded on 11 February 2007 satisfied the aforementioned criteria.    

\subsection{Reduction \& Calibration}

 For consistency, the reduction and calibration of all images was repeated.  The data were reduced, using standard object-minus-sky (A-B) subtraction and flat-fielding.  Individual images from dithering were coadded to remove detector artefacts.  In order to enable a robust comparison of all images, we blurred the 8-meter Gemini/T-ReCS and Subaru/COMICS data according to the diffraction-limited spatial resolution achieved on a 3-meter primary ($\sim$0.7``) and resampled the images at the 0.26'' pixel scale of IRTF-MIRSI.
 
 A limb-fitting procedure was used to assign latitudes,  longitudes and viewing geometries to each pixel on the disk of Jupiter.  Absolute calibration was performed by deriving the central meridian radiance vs latitude profile from the image as well as mid-infrared spectra recorded by Voyager-IRIS and Cassini-CIRS and scaling the former profile to match the latter.   This approach is described in greater detail in \citet{fletcher_2009}.  While this does make the images insensitive to long-term changes, our analysis calculates the ratio of radiance between the auroral hotspot and a lower-latitude zonal mean (see further details in Section \ref{sec:temporal}), in which case the absolute calibration scale factor cancels out. The noise-equivalent spectral radiances (NESR) of the images were calculated by determining the standard deviation of sky pixels more than 5” away from Jupiter’s limb.  

 \begin{figure*}[t!]
\begin{center}
\includegraphics[width=0.8\textwidth]{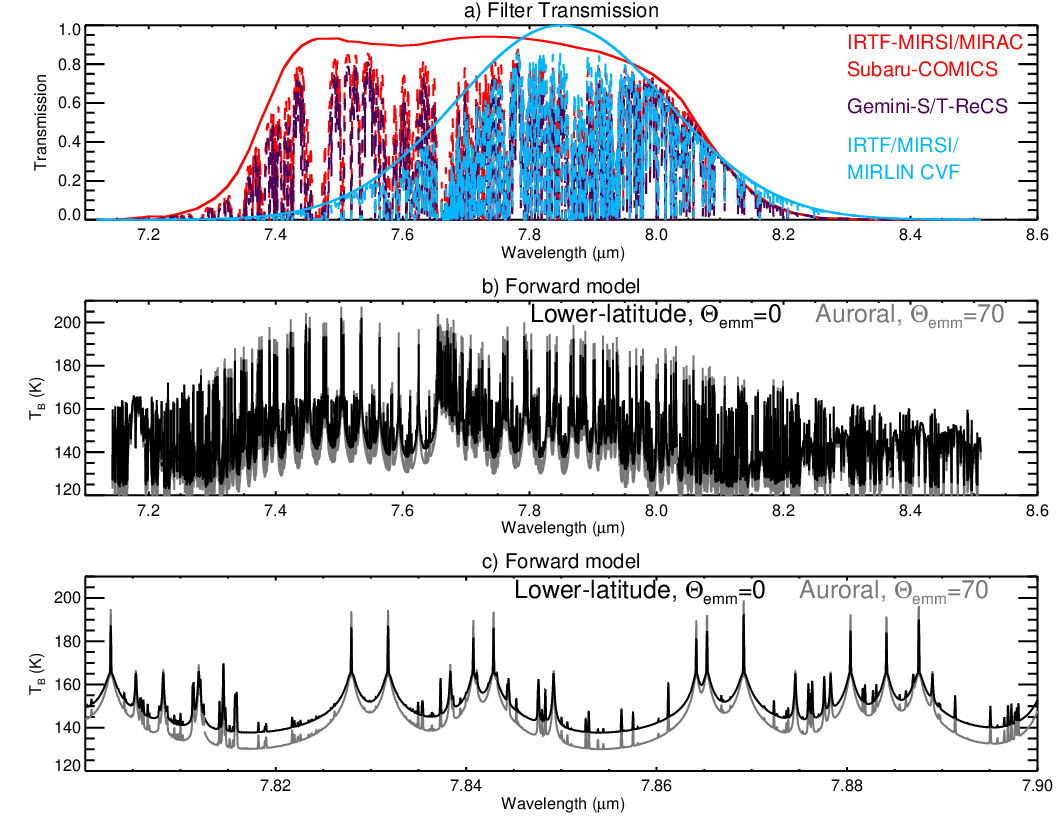}
\caption{a) The transmission filter functions for the instruments used in this work according to the legend.  Solid lines represent the filter transmission, dashed lines are the filter transmission convolved with the telluric transmission spectrum.  Red represents MIRSI/MIRAC on the IRTF and COMICS on Subaru (all at the summit of Mauna Kea), blue represents the CVF setting of IRTF MIRSI/MIRLIN and magenta represents Gemini-South/T-ReCS on Cerro Pachon.  b) forward model spectra of a lower-latitude at nadir (black) and the auroral hotspot at an emission angle of 60$^\circ$ (grey), c) as in b) but over the 7.80 - 7.90 $\upmu$m range such that invidividual lines can be seen.}
\label{fig:plot_filters}
\end{center}
\end{figure*}

\subsection{Transmission Functions}

The filter transmission functions for MIRSI at 7.7 $\upmu$m, COMICS at 7.8 $\upmu$m, MIRAC and T-ReCS at 7.9 $\upmu$m use an identical filter: each observatory has labelled the central filter wavelength differently depending on whether they used the effective wavelength before or after convolution with fore-optics and/or telluric transmission.   A subset of MIRAC images and all MIRLIN images were recorded using a 5\%-wide circular variable filter (CVF) centered at 7.85 $\upmu$m.   Figure \ref{fig:plot_filters}a compares the filter transmission functions of the observations used in this work with spectra of Jupiter, which were calculated as detailed in the next Section. 
 
\section{Analysis}

\subsection{Radiative Transfer Simulations}\label{sec:rad}

Our dataset includes images recorded over different filter bandpasses, observatory altitudes and relative Earth-Jupiter velocities.  We performed radiative transfer modelling in order to quantify how these parameters would affect absolute, observed broadband radiances, and how such variations could be corrected.

We used the NEMESIS software  \citep{irwin_2008} in order to perform radiative transfer forward modelling.  Calculations were performed using the line-by-line treatment over 7.0 to 8.5 $\upmu$m, at a spectral resolving power $\lambda / d\lambda$ = 130000, which sufficiently resolves individual weak and strong emission lines of CH$_4$.  The sources of spectroscopic line data for the H$_2$ S(1) quadrupole line feature, NH$_3$, PH$_3$, CH$_4$, CH$_3$D, $^{13}$CH$_4$, C$_2$H$_2$, C$_2$H$_4$ and C$_2$H$_6$ were adopted from \citet{fletcher_poles_2018}.  The vertical profiles of H$_2$, He, NH$_3$, PH$_3$ were adopted from \citet{sinclair_2018a}.  The vertical profiles of CH$_4$ and its photochemical by-products were taken from a photochemical model based on \citet{moses_2017} but outputting the model results at 60$^\circ$N and assuming an eddy diffusion coefficient profile that results in a CH$_4$ homopause of 7.3 nbar.   This is to be consistent with the findings of \citet{sinclair_2020b}, which demonstrated the CH$_4$ homopause altitude is higher inside Jupiter's northern auroral oval region compared to elsewhere on the planet.  We note that the NEMESIS radiative transfer code assumes local thermodynamic equilibrium (LTE) while the atmosphere is expected to deviate from LTE conditions at pressures lower than 0.1 mbar (e.g. \citealt{appleby_1990,sinclair_2020b}).  Parameterizing non-LTE (NLTE) will be the subject of future work. 

Calculations of synthetic spectra and vertical functional derivative calculations, which describes the relative contribution of each atmospheric level to the total observed radiance, were performed twice. First, at a nadir emission angle using the temperature profile adopted in the \citet{moses_2017} photochemical model. This uses results derived from the Infrared Space Observatory's Short Wave Spectrometer, ISO-SWS (\citealt{lellouch_2001} and references therein) at pressures greater than 1 mbar and Galileo Atmospheric Structure Instrument, ASI \citep{seiff_1998} measurements at pressures lower than 1 mbar.   We use this calculation as representative of the spectrum and atmosphere at a lower latitude.  Second, the calculation was performed at an emission angle of 60$^\circ$ using the temperature profile retrieved by \citet{sinclair_2020b} using a mean of IRTF-TEXES measurements recorded on August 20, 2019 at 68$^\circ$N at longitudes magnetsopherically-poleward of and including the main auroral oval \citep{bonfond_2017}.  This calculation is representative of the geometry and thermal structure of the auroral hotspot.

Figure \ref{fig:plot_filters}b-c shows both synthetic spectra.  For the auroral hotspot, emission from the CH$_4$ line cores is higher due to the upper-stratospheric heating and limb brightening whereas radiances in the continua are limb-darkened and dimmer than those of the nadir, lower-latitude spectrum.  Both spectra, $I(\lambda, v_r)$, where $I$ is the radiance, were doppler-shifted to simulate Jupiter being observed by a relative Earth-Jupiter velocity of $v_r$.  We performed this calculation using a range of $v_r$ from -28 to +28 km/s, in increments of 1 km/s,  in order to capture the range of relative Earth-Jupiter velocities of the data presented in this work (Table \ref{tab:obs}).  The doppler-shifted spectrum was then convolved with the filter response function, $R(\lambda)$, the telluric transmission spectrum $\tau(\lambda)$  and then integrated over the wavelength range ($\lambda_1$ - $\lambda_2$) of the filter bandpass, as shown in Equation \ref{eq:photometry1}.  The telluric transmission spectrum was calculated using the ATRAN\footnote{https://atran.arc.nasa.gov/cgi-bin/atran/atran.cgi} software \citep{lord_1992} assuming a zenith angle of 45$^\circ$, a precipitable water vapor of 1 mm and at altitudes of 8900 (Cerro Pachon) and 13800 feet (Mauna Kea).  A similar calculation was then performed for the 2-dimensional vertical functional derivative image $d I/d T (\lambda,p, v_r)$, as shown in Equation \ref{eq:photometry2}.  

\begin{equation}\label{eq:photometry1}
 \bar{I} (p, v_r) = \frac{\int_{\lambda_1}^{\lambda_2} I(\lambda,p, v_r) * R(\lambda) * \tau(\lambda)  d\lambda} {\int_{\lambda_1}^{\lambda_2}  R(\lambda) * \tau(\lambda) d\lambda}
\end{equation}

\begin{equation}\label{eq:photometry2}
\frac{d \bar{I}}{d T} (p, v_r) = \frac{\int_{\lambda_1}^{\lambda_2} \frac{d I}{d T} (\lambda,p, v_r) * R(\lambda) * \tau(\lambda)  d\lambda} {\int_{\lambda_1}^{\lambda_2}  R(\lambda) * \tau(\lambda) d\lambda}
\end{equation}

\begin{figure}[t]
\begin{center}
\includegraphics[width=0.5\textwidth]{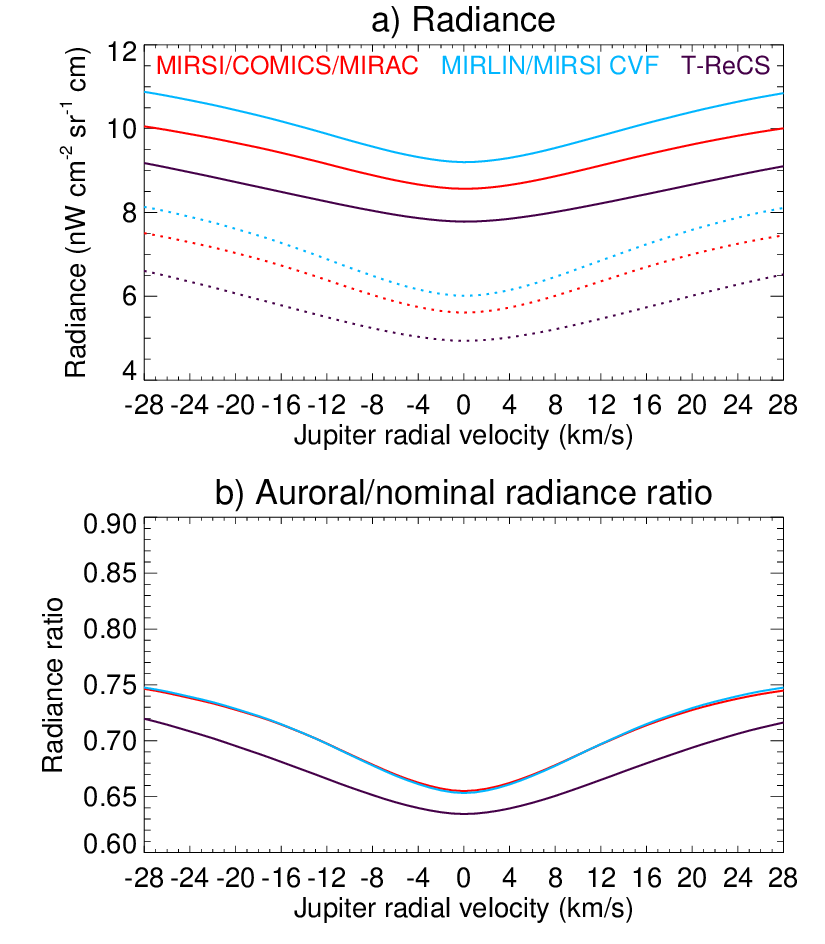}
\caption{a) Variations in the total observed radiance as a function of the relative Earth-Jupiter velocity for MIRSI at 7.7 $\upmu$m/COMICS at 7.8 $\upmu$m/MIRAC at 7.9 $\upmu$m (red), MIRAC/MIRSI CVF at 7.85 $\upmu$m (light blue) and T-ReCS on Gemini-South (purple).   Solid lines show radiance assuming a low latitude temperature profile and a nadir emission angle, dotted lines show radiance assuming a typical auroral hotspot temperature profile and an emission angle of 60$^\circ$. b) The ratio of the auroral-to-nominal radiance.}
\label{fig:plot_rad_vs_vrad}
\end{center}
\end{figure}

Figure \ref{fig:plot_rad_vs_vrad}a shows the variations in absolute, broadband radiance at both the lower-latitude and auroral hotspot locations for each instrument/observatory and as a function of $v_r$.  At a given $v_r$, the same spectrum would produce a range of absolute, convolved radiances due to differences in the filter bandpasses and the altitudes of each instrument/observatory.  By calculating the ratio of radiance between the auroral hotspot and a lower latitude location, such differences are effectively removed for MIRSI/MIRAC/COMICS and the CVF MIRSI/MIRLIN filters, as demonstrated in Figure \ref{fig:plot_rad_vs_vrad}b.  For T-ReCS on Gemini-South, the radiance ratio vs Jupiter radial velocity curve is offset by less than 5\% with respect to the Mauna Kea instruments.  Ultimately, our analysis adopts only one Gemini/T-ReCS measurement and the propagated uncertainty on the radiance ratio from random noise in the image and the standard deviation on pixels averaged is $\sim$10\%.  We therefore consider the offset in radiance ratio of the Gemini/T-ReCS measurement with respect to the Mauna Kea measurements to already be captured within this uncertainty. 

\begin{figure*}[t]
\begin{center}
\includegraphics[width=0.66\textwidth]{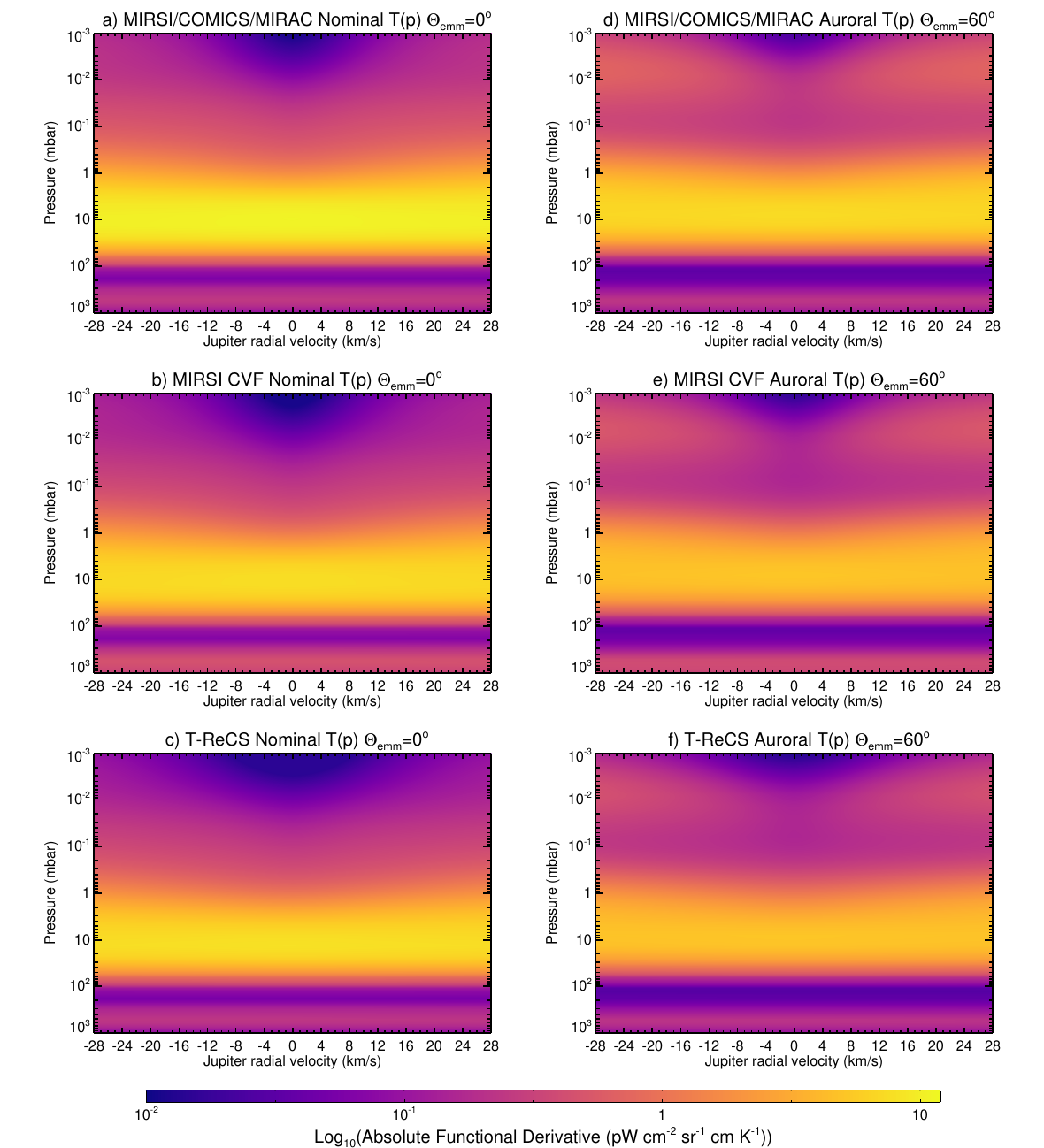}
\caption{The logarithmic, vertical functional derivatives with respect to temperature, which describes the contribution of each atmospheric level to the total observed radiance at the top of the atmosphere.  The left column shows the functional derivatives assuming a nadir emission angle and a lower-latitude temperature profile for a) MIRSI at 7.7 $\upmu$m/COMICS at 7.8 $\upmu$m/MIRAC at 7.9 $\upmu$m, b) MIRSI/MIRSI CVF, and c) T-ReCS on Gemini-South.  The right column shows similar results assuming a typical auroral hotspot temperature profile and calculated at an emission angle of 60$^\circ$.  }
\label{fig:plot_contr_fns}
\end{center}
\end{figure*}

For a given spectrum, higher, convolved radiances are recorded when Jupiter is observed at higher $v_r$, since jovian CH$_4$ emission lines are doppler-shifted out of the telluric CH$_4$ lines to a greater extent.  Higher relative velocities also better sample the strong cores of the CH$_4$ emission lines, which in turn sound higher and warmer altitudes in Jupiter's atmosphere.  This is particularly true for Jupiter's auroral hotspot due to its high latitude/emission angle, and the associated limb-brightening effect, as well as the upper stratospheric heating present in this region.  This is demonstrated in Figure \ref{fig:plot_contr_fns}, which shows the vertical functional derivatives for each instrument for both a nadir, lower-latitude and a high emission angle observation of the auroral hotspot.  A consequence of this is that calculating the ratio in radiance between the auroral hotspot and a lower latitude does not remove the apparent variability in brightness due to changes in $v_r$ alone, as shown in Figure \ref{fig:plot_rad_vs_vrad}b.  We address this in two ways.  First, we use the curves shown in Figure \ref{fig:plot_rad_vs_vrad}a-b to derive a correction for radial velocity, which we apply to the data as detailed further in Section \ref{sec:temporal}.  Second, in Section \ref{sec:temporal}, we demonstrate that there is negligible correlation of the (uncorrected) radiance of the auroral hotspot with Jupiter's relative velocity at the time of measurements.

\subsection{Searching for temporal patterns}\label{sec:temporal}

For each image, a centre-to-limb correction was applied in order to correct for limb brightening and foreshortening.  The mean radiance between 60 - 75 $^\circ$N and 160 - 200 $^\circ$W was calculated, which is a latitude/longitude range that captures the emissions poleward of the main oval.  The standard deviation of the mean, and the noise-equivalent radiance scaled by n$^{-0.5}$, where n is the number of diffraction-resolved pixels, were calculated and the larger of the two was adopted as the uncertainty on the average radiance.  Similarly, the average radiance was calculated between 50$^\circ$S - 50$^\circ$N at all longitudes captured in the image at $\upmu > $ 0.2.   The assumption is that the mean radiance between 50$^\circ$S and 50$^\circ$N should not exhibit any net time variability since seasonal variations in each hemisphere should cancel out and smaller-scale variable or transient features, such as the equatorial oscillation (e.g. \citealt{giles_2020,antunano_2021,orton_2023}), stratospheric waves (e.g. \citealt{fletcher_2017}) should have negligible effect on the mean radiance calculated over a large area.  

Using the radiance vs $v_r$ curves derived in Figure \ref{fig:plot_rad_vs_vrad}, a radial-velocity correction was applied to both sets of radiances.  We then calculated the ratio of the auroral to lower-latitude radiance in these locations, henceforth described as the `Relative Poleward Radiance' or RPR.  This effectively cancels out any variations in absolute radiance due to differences in filters/observatory altitudes as well as absolute calibration.  On nights where more than one image satisfied the aforementioned criteria, a mean RPR was calculated.  We also performed the same calculation for radiances without the radial velocity correction to demonstrate that Jupiter's radial velocity is not the dominant driver of the northern auroral hotspot's variability (see Section \ref{sec:rule_out}).

\subsubsection{Ruling out relative velocity}\label{sec:rule_out}

\begin{figure*}[t!]
\centering
\includegraphics[width=0.5\textwidth]{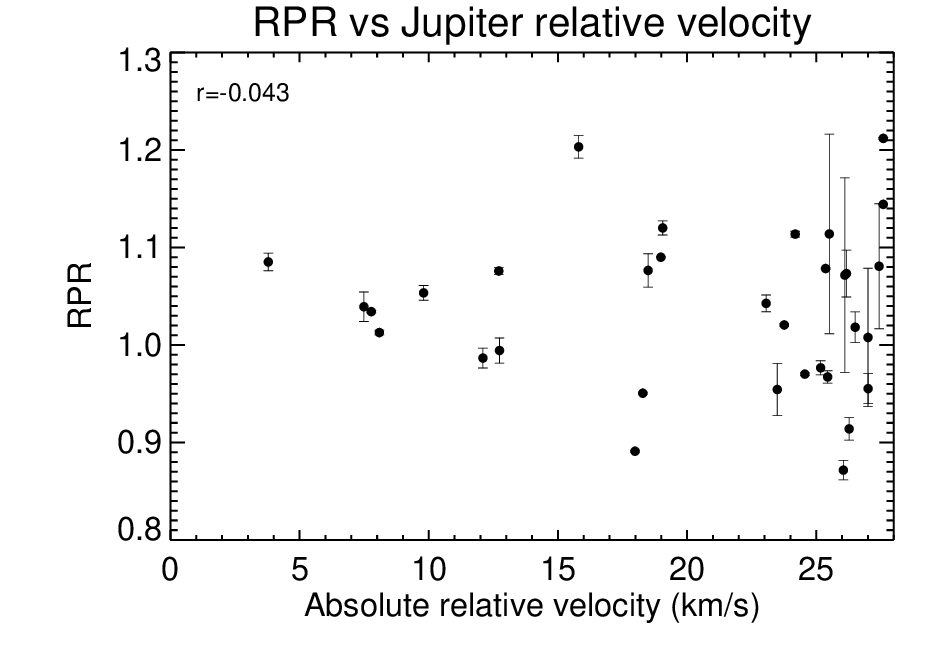}
\caption{The relative poleward radiance (RPR) of Jupiter's auroral hotspot compared to the absolute Earth-Jupiter relative velocity.  The correlation coefficient is shown in the top-left.}
\label{fig:test_vs_vrad}
\end{figure*}

First, we compared the (uncorrected) RPR with Jupiter's relative velocity to confirm that the variability of the auroral hotspot is not an artefact of varying Doppler shift. We adopted the observer range rate as a function of time with a step of 1 day using the JPL Horizons system\footnote{https://ssd.jpl.nasa.gov/horizons}, which takes into account the light-time abberation.  Our analysis focuses on observations recorded when the northern auroral hotspot is close or near Jupiter's central meridian, at which time any rotational or horizontal-wind component is negligible compared to the relative velocity between Earth and Jupiter.

Figure \ref{fig:test_vs_vrad} shows a scatter diagram of the (uncorrected) relative poleward radiance as a function of Jupiter's absolute relative velocity.   At first glance, there may appear to be more variabiltiy of the RPR at higher, absolute radial velocities.  However, we consider this a measurement bias since a large proportion of measurements were performed at higher relative velocities.  We derive a correlation coefficient of -0.04 and therefore a negligible correlation of the RPR with Jupiter's relative velocity.  We therefore conclude that the variability of the auroral hotspot is not a result of Jupiter's varying radial velocity with respect to Earth.

\subsubsection{Longer-term solar insolation}\label{sec:subsolar}

\begin{figure*}[t]
\begin{center}
\includegraphics[width=0.7\textwidth]{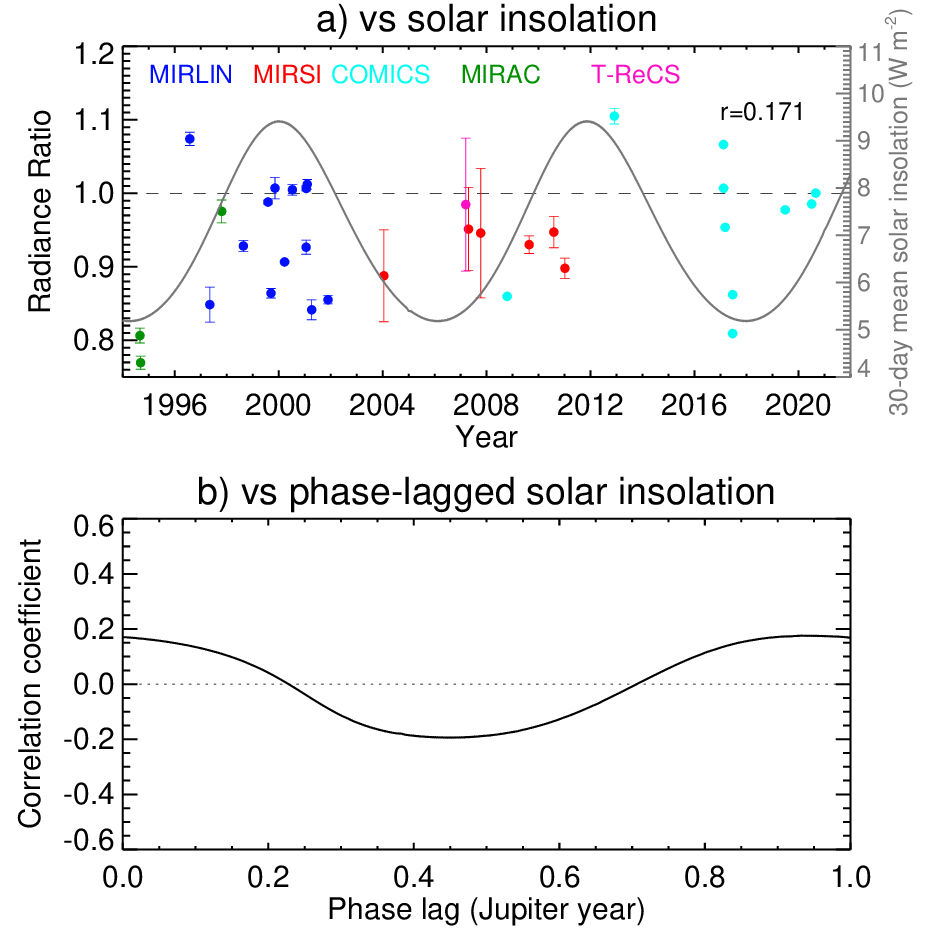}
\caption{a) The relative poleward radiance (RPR) of the northern auroral hotspot as a function of time, according to the left-hand axis.  Points are colored according to the observatory/instrument used as shown in the legend.  The variability is compared to the 30-day mean solar insolation at 65$^\circ$N, 180$^\circ$W according to the right-hand axis. The correlation coefficient of the RPR with solar insolation is shown in the top-right of the panel. b) The correlation coefficient of the RPR vs phase-lagged solar insolation.}
\label{fig:plot_solar}
\end{center}
\end{figure*}

Figure \ref{fig:plot_solar}a shows the variability of the northern auroral hotspot as a function of time.  We find the CH$_4$ emissions of Jupiter's northern auroral region exhibit significant temporal variability over a range of $\sim$37\%. We reproduce the result described in \citet{sinclair_2019a} and \citet{sinclair_2023b}, which demonstrated daily variability of the northern auroral CH$_4$ emissions.  Variability is observed over a range of timescales from days to years.  However, we suspect that the northern auroral hotspot in CH$_4$ emissions exhibits daily variability in response to magnetospheric and solar-wind events and will appear to exhibit erratic variability when sampled irregularly/sparsely by the data in this work.

First, we test whether the variability of the northern auroral hotspot is linked to the varying solar insolation received at Jupiter's high-northern latitudes over the course of a Jupiter year (12 Earth years).  JPL Horizons was used to compute the subsolar latitude and heliocentric range of Jupiter from 1975 to 2030 in steps of 2 hours or 1/5th of a Jupiter rotation.  For each timestep, the solar zenith angle and solar insolation at 65$^\circ$N, 180$^\circ$W were calculated.  The result was then smoothed using a triangular function with a width of 30 (Earth) days thereby removing diurnal variations.  Figure \ref{fig:plot_solar}a compares the variability of the RPR with the 30-day or monthly-mean solar insolation.  Both visually and quantitatively by calculation of the correlation coefficient (r = 0.176), we find there is negligible correlation between the radiance of the northern auroral region with instantaneous solar insolation. 
 
Given that the observations predominantly sound the 10- to 1-mbar level (Section \ref{fig:plot_contr_fns}, where the thermal inertia timescale of the atmosphere is on the order of several Earth years \citep{zhang_2013a}, we next explored whether the radiance of the auroral hotspot varied in accordance with solar insolation but with a phase lag.  Figure \ref{fig:plot_solar}b shows the correlation coefficient of the RPR with sub-solar latitude with a phase lag from 0 to 1 Jovian year.  As shown, there is only marginal change in the correlation coefficient when a phase lag is imposed on the solar insolation.  The maximum and minimum correlation coefficients of 0.18 and -0.19 occur at phase lags of 0.93 and 0.45 Jupiter years, respectively, neither of which are high enough to conclude a correlation exists. 

\subsubsection{Longer-term solar activity}\label{sec:sunspot}

\begin{figure*}[t]
\begin{center}
\includegraphics[width=0.7\textwidth]{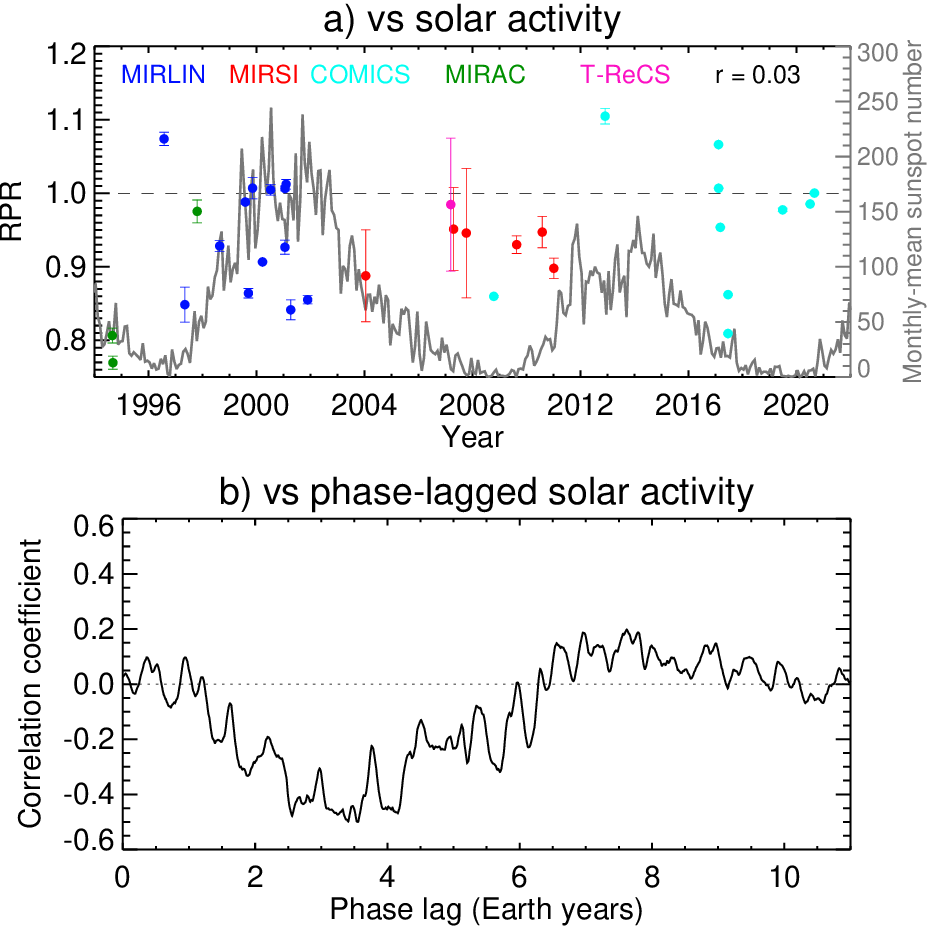}
\caption{a) The relative poleward radiance (RPR) of the northern auroral hotspot as a function of time, according to the left-hand axis.  Points are colored according to the observatory/instrument used as shown in the legend.  The variability is compared to the monthly-mean sunspot number as a metric of solar activity according to the right-hand axis. The correlation coefficient of the RPR with sunspot number is shown in the top-right of the panel. b) The correlation coefficient of the RPR vs phase-lagged solar insolation.}
\label{fig:plot_solarcycle}
\end{center}
\end{figure*}

Second, we explored whether the radiance of the northern auroral hotspot exhibited correlation with the 11-year solar cycle.  Figure \ref{fig:plot_solarcycle}a compares the variability of the RPR with the monthly mean sunspot number, as taken from NOAA\footnote{https://solarscience.msfc.nasa.gov/greenwch/SN\_m\_tot\_V2.0.txt}.  As  in Section \ref{sec:subsolar}, we also tested whether a significant correlation coefficient existed with the phase-lagged monthly mean sunspot number from 0 to 11 years.  There results are shown in Figure \ref{fig:plot_solarcycle}b.

With zero phase lag, we find a near-zero (r $<$ 0.05) correlation of the variability of the auroral hotspot with the monthly-mean sunspot number.  We find a maximum correlation of 0.19 when introducing a phase lag of 7.6 (Earth) years, which we also consider negligible.  A substantially larger minimum correlation coefficient (r = -0.50) is found when adopting a phase lag of $\sim$3.6 Earth years.  A negative correlation of the auroral hotspot radiance with solar activity would be physically-implausible and so we reject this as a sampling artefact.

\begin{figure*}[t]
\begin{center}
\includegraphics[width=0.6\textwidth]{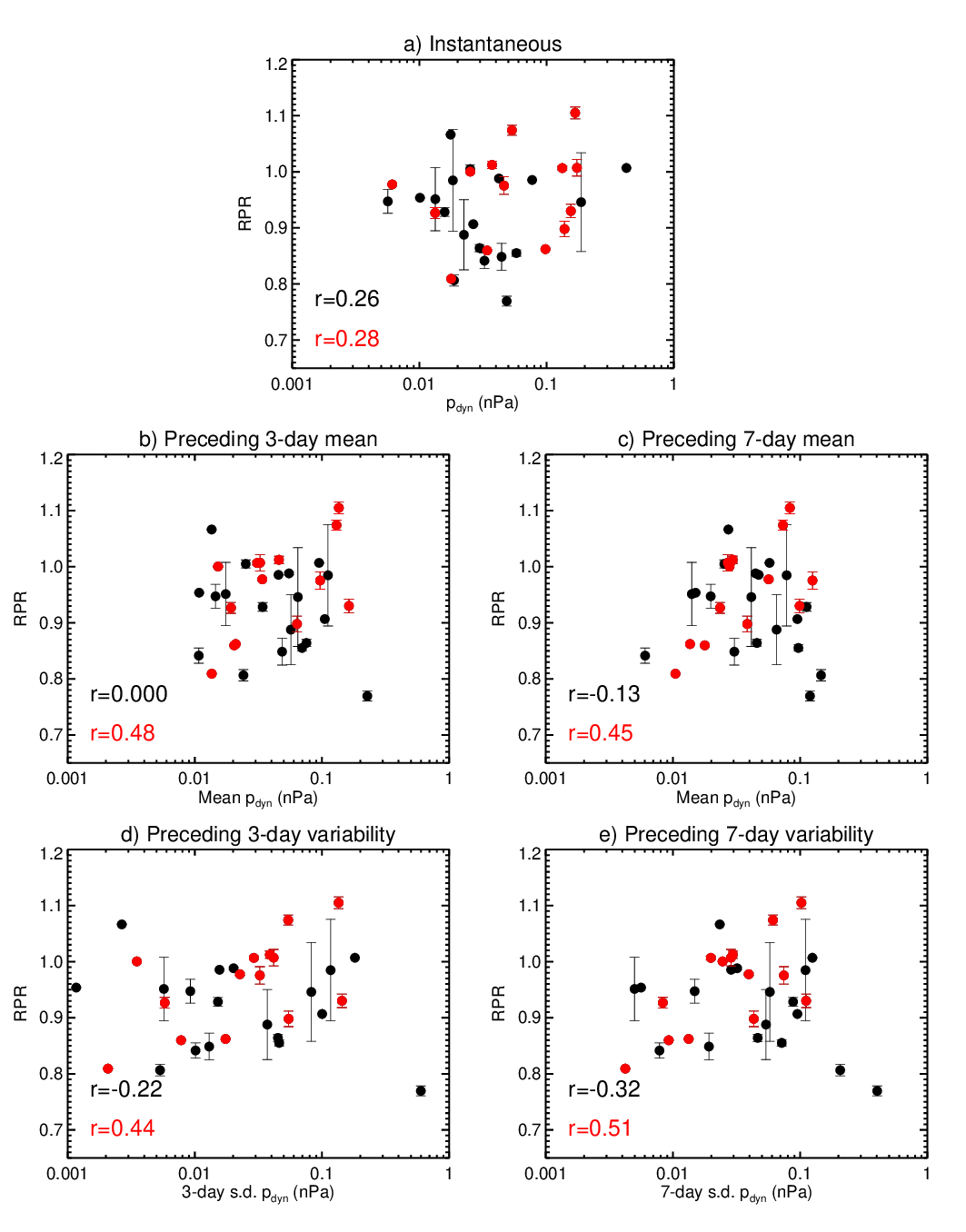}
\caption{Correlations of the relative poleward radiance (RPR) with a) the instantaneous solar wind dynamical pressure (p$_{dyn}$ (nPa)), the mean p$_{dyn}$ in the preceding b) 3 days, c) 7 days, and the standard deviation of p$_{dyn}$ in the preceding d) 3 days and e) 7 days.  The correlation coefficient, r,  in each case is given in bottom, left of each panel in black. Data corresponding to times when Jupiter was within 60$^\circ$ of opposition are colored in red and the correlation coefficient is also shown.  }
\label{fig:solar_wind}
\end{center}
\end{figure*}

\subsection{Short-term solar wind conditions}\label{sec:wind}

Finally, we explored whether the radiance of the northern auroral hotspot exhibited correlation with short-term solar wind conditions at Jupiter.  OMNI-measured solar wind conditions at Earth \citep{thatcher_2011} and the \citet{tao_2005} solar wind propagation model were used to predict the solar wind dynamical pressure, p$_{dyn}$ impinging on Jupiter's magnetosphere from 1994 to 2021.

First, we compared the RPR with the instantaneous solar wind dynamical pressure taking into account the light-time abberation, and found a correlation of r = 0.26, as shown in Figure \ref{fig:solar_wind}a.  As noted in previous work (e.g. \citealt{sinclair_2019b}), solar wind propagation models are 1-dimensional and therefore most accurate when the Earth-Sun-Jupiter angle is small i.e. when Jupiter is close to opposition.  \citet{zieger_2008} quantified the uncertainty on a similar solar wind propagation model and estimated a 12-hour uncertainty on timing and 38\% error on magnitude when Jupiter was within 60$^\circ$ of opposition.  In using only the 14 measurements recorded when Jupiter was within 60$^\circ$ of opposition (shown in red in Figure \ref{fig:solar_wind}a), the correlation coefficient was re-calculated and we found a negligible increase in correlation to r = 0.28.  

Similar calculations were repeated using the mean and standard deviation of solar wind dynamical pressure in the 3 and 7 days preceding the Jupiter images in order to test whether there was a correlation with active and variable periods of solar wind conditions.  We find there is negligible correlation when performing the calculation using all 33 measurements.  However, in using only the 14 measurements recorded when Jupiter was within 60$^\circ$ of opposition, when solar-wind propagation model results are more certain, we find moderate correlations of the relative poleward radiance with all four, aforementioned metrics of solar wind activity and variability.  For example, the RPR exhibits a r = 0.51 correlation with the standard deviation of the solar wind dynamical pressure in the 7 days preceding the Jupiter images.

\section{Discussion}

We collated an extensive record of 7.7 - 7.9 $\upmu$m images of Jupiter from Earth-based telescopes over a period of three decades.  These images allowed us to quantify the timescales and magnitude over which the northern auroral hotspot CH$_4$ emissions varied.

Figure \ref{fig:plot_solar} shows the relative poleward radiance or RPR of the northern auroral hotspot over the 1994 to 2021 time period.  The RPR is the ratio in radiance of the northern auroral hotspot with a lower-latitude zonal mean, which removes/minimizes uncertainty due to absolute calibration, differences in filter bandpasses and observatory altitudes.   We find the CH$_4$ emissions of Jupiter's northern auroral hotspot exhibit significant temporal variability over a range of $\sim$37\%.   This reproduces the result described in \citet{sinclair_2019a} and \citet{sinclair_2023b}, which demonstrated daily variability of the northern auroral CH$_4$ emissions using different datasets.  Variability is observed over a range of timescales from days to years.  We suspect that the CH$_4$ emission of the northern auroral hotspot exhibits daily variability in response to magnetospheric and solar-wind events and will appear to exhibit erratic variability when sampled irregularly/sparsely.  

First, we tested whether the variability of the northern auroral hotspot could be explained by the varying solar insolation at Jupiter's high-northern latitudes.  We found there was negligible correlation with both the instantaneous solar insolation as well the phase-lagged solar insolation. This allows us to explicitly rule out the hypothesis we suggested in previous work (e.g. \citealt{sinclair_2017a,sinclair_2017b}) that shortwave solar heating of haze particles produced by the auroral chemistry (e.g. \citealt{wong_2000,friedson_2002,wong_2003}) is the dominant source of 1-mbar auroral-related heating.  If this source of heating were significant, given the $\sim$3 year thermal inertia/radiative lifetimes at this altitude (e.g. \citealt{zhang_2013a}), 1-mbar temperatures would vary in accordance with seasonally-changing solar insolation due to Jupiter's $\sim$3$^\circ$ axial tilt.  The majority of the broadband 7.7-7.9$\upmu$m CH$_4$ emissions originate from close to this pressure level (Figure \ref{fig:plot_contr_fns}).  Thus, such hypothetical 1-mbar temperature variations would be observed as a long-term modulation of the 7.7-7.9 CH$_4$ emissions in accordance with solar insolation, which is not observed.  While solar heating of haze particles likely has a non-negligible contribution to the radiative balance of the auroral stratosphere, it cannot be the dominant source of auroral heating at 1 mbar.

Second, we also tested whether the radiance of the northern auroral hotspot varied in accordance with the 11-year solar cycle. We also found negligible correlation of the variability of the hotspot with the monthly-mean sunspot number as a metric of solar activity over the 11-year solar cycle.  We also found negligible correlation when introducing a phase lag to the monthly mean sunspot number.   We therefore dismiss the hypothesis that the CH$_4$ emissions of the auroral hotspot exhibit a longer-term variability associated with the 11-year solar cycle.  The temporal sampling of our data does not allow us to confidently conclude whether or not the CH$_4$ emissions of the auroral hotspot exhibit more variability during periods of solar maxima, as has been interpreted for C$_2$H$_6$ emissions \citep{kostiuk_agu_2016}.   For example, we do see high variability in the 1999-2002 time period in proximity to the maximum of Solar cycle 23, however, we do not have a similar sampling during periods of solar minima to say whether the auroral hotspot is as variable, or more/less variable.   

Third, we tested whether the auroral hotspot exhibited correlation with short-term solar wind conditions.  Solar-wind conditions measured at Earth using OMNI \citep{thatcher_2011} and the \citet{tao_2005} solar wind propagation model was used to predict the solar wind conditions impinging on Jupiter's magnetosphere.  Such solar wind propagation models are expected to have the lowest uncertainty when Jupiter is within 60$^\circ$ of opposition, i.e. when the Earth-Jupiter-Sun angle is small.  We therefore tested the correlation of the radiance of the auroral hotspot with solar wind conditions using: 1) all available Jupiter images, 2) only those recorded when Jupiter was within 60$^\circ$ of opposition - see Section \ref{sec:wind}.  Here, we only discuss the results of the latter.  The radiances of the northern auroral hotspot exhibit negligible correlation with the instantaneous solar wind dynamical pressure, p$_{dyn}$ at Jupiter (taking into account the light-time abberation).  However, we find moderate correlations of the auroral hotspot with the mean p$_{dyn}$ and the standard deviation of p$_{dyn}$ in the days preceding the Jupiter images.  For example, we find the relative poleward radiance exhibits an r = 0.45 correlation with the preceding 7-day mean p$_{dyn}$ and an r = 0.51 correlation with the standard deviation on the 7-day mean.  This demonstrates that for a subset of the times the northern auroral hotspot exhibited enhanced CH$_4$ emissions, there were strong and variable solar wind conditions present in the preceding days/week.  The fact the correlation is not higher likely denotes that either internal magnetospheric processes, independent of the solar wind (e.g. Io's output) can also produce a brightening of the mid-infrared auroral hotspot and/or only a subset of solar wind enhancements lead to brightening of the auroral emissions.   In a similar analysis of Jupiter's ultraviolet auroral emissions using HISAKI-EXCEED, \citet{kita_2016} noted correlations of a similar magnitude ($\sim$0.44) between the total auroral power and the preceding variation in solar wind dynamical pressure.  However, their spatially-coarse observations could not distinguish whether the variations in auroral power were arising from the main emission, the poleward emissions or outer emissions.   Using the same dataset but focusing on the emissions from the Io Plasma torus (IPT) at 5 - 7 R$_J$, \citet{murakami_2016} also demonstrated that dawn-dusk asymmetries were amplified during periods of solar wind compressions.  This demonstrated that solar wind variations could ultimately perturb Jupiter's inner magnetosphere.

Using observations recorded in 2013, 2018 and 2020, \citet{kim_2020, kim_2022} concluded there was negligible correlation of the 3-$\upmu$m auroral CH$_4$ emissions with external solar wind conditions at Jupiter, in contrast to the moderate correlations of the 8-$\upmu$m emissions reported in this work.  One possible source of this difference is that the 3-$\upmu$m emissions predominantly sound microbar pressures while the 8-$\upmu$m broadband emissions predominantly sound millibar pressures, and therefore exhibit contrasting temporal behaviours due to differences in the thermal inertia and radiative lifetimes \citep{zhang_2013a}.  Another possible reason is that the observations analyzed by \citet{kim_2020} and \citet{kim_2022} by coincidence captured brightenings of the CH$_4$ emissions in response to internal magnetospheric processes without a solar wind perturbation or events where solar wind variations led to no apparent change in CH$_4$ emissions.  Further observations of Jupiter's auroral CH$_4$ emissions at 3 $\upmu$m and 8 $\upmu$m may better highlight similarities and differences in their temporal behaviour. 

As demonstrated in Figure \ref{fig:plot_contr_fns}, broadband 7.7 - 7.9 $\upmu$m imaging of Jupiter's northern auroral hotspot predominantly sounds the lower stratosphere ($\sim$1 mbar).  The modulation of the northern auroral hotspot's 7.7 to 7.9 $\upmu$m CH$_4$ emission by a subset of solar wind events suggests that variable solar wind conditions and their perturbing effects on the magnetosphere can ultimately modulate 1-mbar temperatures on timescales of several days.  Particle precipitation modelling \citep{ozak_2010,gustin_2016,houston_2018} and \textit{in-situ} measurements by the Juno spacecraft (e.g. \citealt{clark_2017,paranicas_2018,mauk_2020}) demonstrate that a negligible flux of magnetospheric particles deposit their energies at pressures deeper than 0.1 mbar.  Thus, we suggest that the modulation of 1-mbar temperatures cannot result from \textit{direct} deposition of energy from the magnetosphere.  Instead, we favor the explanation first suggested in \citet{cavalie_2021}, who observed counterrotating jets at $\sim$0.1 mbar, coincident with the southern auroral oval and inferred that a similar circulation is likely present at the northern auroral oval too.  They suggested that ion-neutral collisions drive the formation of the vortex, and may possibly be an extension of a similar counterrotating electrojet predicted and observed at thermospheric altitudes (e.g. \citealt{achilleos_2001,johnson_2018}).  This could produce atmospheric subsidence poleward of the main oval and therefore adiabatic heating at pressures higher than 0.1 mbar.  Magnetospheric dynamics, driven by both internal processes (e.g. Io volcanism) or by solar wind compressions, could perhaps accelerate the vortex, thereby increasing the magnitude of 1-mbar heating.  The vortex would trap warm gas augmenting the thermal contrast with regions outside of the auroral hotspot.  Periods of quiescent, solar wind conditions could perhaps allow the vortex to decelerate, even dissipate, which would stop adiabatic heating as a heat source and also enable warm gas to be advected/diffused away from the hotspot.  However, time-dependent circulation modelling would be required to determine whether this proposed mechanism can produce the magnitude of heating and variability observed of the auroral hotspot.

\section{Conclusions}

We present a study of the long term variability of Jupiter's mid-infrared CH$_4$ auroral emissions.  7.7-7.9 $\upmu$m images of Jupiter recorded  by NASA's Infrared Telescope Facility, Subaru and Gemini-South over the last three decades were collated.  A ratio of the radiance of the poleward northern auroral emissions with a lower-latitude zonal-mean, henceforth `Relative Poleward Radiance' or RPR was calculated from each image, and compared to quantify the magnitude and timescales over which auroral hotspot varies.  We find the RPR exhibits erratic, variable behaviour on a range of timescales.  The variability of the RPR exhibits a weak (r $<$ 0.2) correlation with both the instantaneous and phase-lagged solar insolation received at Jupiter's high-northern latitudes.  This rules out the hypothesis suggested in previous work (e.g. \citealt{sinclair_2017a,sinclair_2018a}) that shortwave solar heating of aurorally-produced haze is the dominant auroral-related heating mechanism in the lower stratosphere.  We also find the variability exhibits negligible (r $<$ 0.18) correlation with both the instantaneous and phase-lagged monthly-mean sunspot number and therefore rule out a long-term variability associated with the solar cycle.   On shorter timescales, we find moderate correlations of the RPR with solar wind conditions at Jupiter in the preceding days before images were recorded.  For example, we find correlations of r = 0.45 and r = 0.51 of the RPR with the mean and standard deviation solar wind dynamical pressure in the preceding 7 days.  The moderate correlation suggests that either: 1) only a subset of solar wind compressions lead to brighter, poleward CH$_4$ emissions and/or 2) a subset of CH$_4$ emission brightening events are driven by internal magnetospheric processes (e.g. Io activity) and independent of solar wind enhancements.

\section{Acknowledgements} 

The research was carried out at the Jet Propulsion Laboratory, California Institute of Technology, under a contract with the National Aeronautics and Space Administration (80NM0018D0004).  The material is based upon work supported by NASA under Grant NNH19ZDA001N issued through the Cassini Data Analysis (CDAP) program. The High Performance Computing resources used in this investigation were provided by funding from the JPL Information and Technology Solutions Directorate.  The IRTF is operated by the University of Hawaii under contract NNH14CK55B with NASA. Co-author Tao acknowledges the support by MEXT/JSPS KAKENHI Grant 19H01948.  Co-author Fletcher was supported by a European Research Council Consolidator Grant (under the European Union’s Horizon 2020 research and innovation programme, grant agreement No 723890) at the University of Leicester. A subset of observations were recorded at the international Gemini Observatory, a program of NSF’s NOIRLab, which is managed by the Association of Universities for Research in Astronomy (AURA) under a cooperative agreement with the National Science Foundation on behalf of the Gemini Observatory partnership: the National Science Foundation (United States), National Research Council (Canada), Agencia Nacional de Investigaci\'{o}n y Desarrollo (Chile), Ministerio de Ciencia, Tecnolog\'{i}a e Innovaci\'{o}n (Argentina), Minist\'{e}rio da Ci\^{e}ncia, Tecnologia, Inova\c{c}\~{o}es e Comunica\c{c}\~{o}es (Brazil), and Korea Astronomy and Space Science Institute (Republic of Korea).  Some data were also recorded at the Subaru Telescope, which is operated by the
National Astronomical Observatory of Japan. A subset of Subaru data were recorded through the Keck-Subaru time exchange programme.  We acknowledge the W. M. Keck Observatory, which is operated as a scientific partnership between California Institute of Technology, the University of California and NASA and supported financially by the W. M. Keck Foundation. We recognize and acknowledge the very important cultural role and reverence that the summit of Maunakea has always had within the indigenous Hawaiian community. We are most fortunate to have the opportunity to conduct observations from this mountain.

\section{Data Availability}

Archiving of all data recorded in this study to the Planetary Data Systems (PDS) node is in progress (at the time of writing).  Until they are available on the PDS, the data will be made available upon request.

\clearpage
\newpage
\appendix
\section{Data tables}
\FloatBarrier
\begin{table*}[!th]
\begin{tabular}{>{\centering\arraybackslash} m{2.5cm} >{\centering\arraybackslash} m{2.0cm}  >{\centering\arraybackslash} m{3.0cm} >{\centering\arraybackslash} m{1.5cm}>{\centering\arraybackslash} m{1.2cm} >{\centering\arraybackslash} m{1.2cm} >{\centering\arraybackslash} m{1.2cm}} 
Date & Time (UTC) & Instrument & Filter & Airmass & CML & v$_{rad}$ \\
(dd-mmm-yyyy) & (hh:mm) & & ($\upmu$m) & & & (km/s) \\
\firsthline 
28-Jul-1994	&	2:51	&	IRTF/MIRAC	&	7.85	&	1.92	&	176	& 26.3	\\
6-Aug-1994	&	5:34	&	IRTF/MIRAC	&	7.85	&	1.19	&	188	& 26.1	\\
6-Dec-1995	&	0:43	&	IRTF/MIRAC	&	7.85	&	1.51	&	183	&	4.2	\\
8-Dec-1995	&	3:46	&	IRTF/MIRAC	&	7.85	&	1.45	&	166	&	3.5	\\
28-Jun-1996	&	12:47	&	IRTF/MIRLIN	&	7.85	&	1.56	&	160	&-4.0	\\
28-Jun-1996	&	13:32	&	IRTF/MIRLIN	&	7.85	&	1.82	&	187	&-4.0	\\
5-Apr-1997	&	19:03	&	IRTF/MIRLIN	&	7.85	&	1.37	&	169	&-23.3	\\
20-Sep-1997	&	6:12	&	IRTF/MIRAC	&	7.85	&	1.36	&	162	&18.3	\\
20-Jul-1998	&	11:39	&	IRTF/MIRLIN	&	7.85	&	1.46	&	206	&-23.2	\\
1-Jul-1999	&	14:39	&	IRTF/MIRLIN	&	7.85	&	1.41	&	164	&-24.1	\\
13-Aug-1999	&	5:57	&	IRTF/MIRLIN	&	7.85	&	1.20	&	200	&-25.2	\\
8-Oct-1999	&	10:22	&	IRTF/MIRLIN	&	7.85	&	1.04	&	156	&-7.7	\\
21-Feb-2000	&	2:17	&	IRTF/MIRLIN	&	7.85	&	1.06	&	169	&23.8	\\
21-Feb-2000	&	3:20	&	IRTF/MIRLIN	&	7.85	&	1.01	&	208	&23.8	\\
9-Jun-2000	&	17:33	&	IRTF/MIRLIN	&	7.85	&	1.41	&	185	&-9.6	\\
20-Dec-2000	&	5:52	&	IRTF/MIRLIN	&	7.85	&	1.64	&	166	&12.0	\\
20-Dec-2000	&	5:55	&	IRTF/MIRLIN	&	7.85	&	1.62	&	168	&12.0	\\
20-Dec-2000	&	5:51	&	IRTF/MIRLIN	&	7.85	&	1.69	&	165	&12.0	\\
20-Dec-2000	&	5:54	&	IRTF/MIRLIN	&	7.85	&	1.63	&	168	&12.0	\\
20-Dec-2000	&	5:55	&	IRTF/MIRLIN	&	7.85	&	1.62	&	168	&12.0	\\
21-Dec-2000	&	11:53	&	IRTF/MIRLIN	&	7.85	&	1.51	&	175	&12.5	\\
21-Dec-2000	&	11:54	&	IRTF/MIRLIN	&	7.85	&	1.51	&	176	&12.5	\\
21-Dec-2000	&	11:54	&	IRTF/MIRLIN	&	7.85	&	1.51	&	176	&12.5	\\
21-Dec-2000	&	11:55	&	IRTF/MIRLIN	&	7.85	&	1.52	&	176	&12.5	\\
21-Dec-2000	&	11:59	&	IRTF/MIRLIN	&	7.85	&	1.55	&	179	&12.5	\\
21-Dec-2000	&	12:00	&	IRTF/MIRLIN	&	7.85	&	1.55	&	179	&12.5	\\
21-Dec-2000	&	12:00	&	IRTF/MIRLIN	&	7.85	&	1.55	&	179	&12.5	\\
21-Dec-2000	&	12:02	&	IRTF/MIRLIN	&	7.85	&	1.57	&	180	&12.5	\\
21-Dec-2000	&	12:05	&	IRTF/MIRLIN	&	7.85	&	1.59	&	182	&12.5	\\
21-Dec-2000	&	12:06	&	IRTF/MIRLIN	&	7.85	&	1.59	&	183	&12.5	\\
21-Dec-2000	&	12:06	&	IRTF/MIRLIN	&	7.85	&	1.59	&	183	&12.5	\\
21-Dec-2000	&	12:07	&	IRTF/MIRLIN	&	7.85	&	1.61	&	183	&12.5	\\
21-Dec-2000	&	12:10	&	IRTF/MIRLIN	&	7.85	&	1.64	&	185	&12.5	\\
21-Dec-2000	&	12:11	&	IRTF/MIRLIN	&	7.85	&	1.64	&	186	&12.5	\\
21-Dec-2000	&	12:11	&	IRTF/MIRLIN	&	7.85	&	1.64	&	186	&12.5	\\
21-Dec-2000	&	12:12	&	IRTF/MIRLIN	&	7.85	&	1.66	&	187	&12.5	\\
5-Jan-2001	&	8:49	&	IRTF/MIRLIN	&	7.85	&	1.07	&	163	&18.9	\\
5-Jan-2001	&	8:52	&	IRTF/MIRLIN	&	7.85	&	1.07	&	164	&18.9	\\
5-Jan-2001	&	8:54	&	IRTF/MIRLIN	&	7.85	&	1.07	&	166	&18.9	\\
5-Jan-2001	&	9:26	&	IRTF/MIRLIN	&	7.85	&	1.13	&	185	&18.9	\\
5-Jan-2001	&	9:24	&	IRTF/MIRLIN	&	7.85	&	1.14	&	184	&18.9	\\
5-Jan-2001	&	9:30	&	IRTF/MIRLIN	&	7.85	&	1.15	&	187	&18.9	\\
\lasthline
\end{tabular}
\caption{The list of 7 - 8 $\upmu$m observations adopted in this study after omitting observations recorded when Jupiter was above an airmass of 2 and when the sub-observer longitude was more than 30$^\circ$ away from 180$^\circ$W.  }
\label{tab:obs}
\end{table*}

\setcounter{table}{0}
\begin{table*}[!th]
\centering  
\begin{tabular}{>{\centering\arraybackslash} m{2.5cm} >{\centering\arraybackslash} m{2.0cm}  >{\centering\arraybackslash} m{3.0cm} >{\centering\arraybackslash} m{1.5cm}>{\centering\arraybackslash} m{1.2cm} >{\centering\arraybackslash} m{1.2cm} >{\centering\arraybackslash} m{1.2cm}} 
Date & Time (UTC)& Instrument & Filter & Airmass & CML & v$_{rad}$ \\
(dd-mmm-yyyy) & (hh:mm) & & ($\upmu$m) & & & (km/s) \\
\firsthline 
7-Mar-2001	&	3:57	&	IRTF/MIRLIN	&	7.85	&	1.01	&	162	&27.0	\\
23-Oct-2001	&	13:00	&	IRTF/MIRLIN	&	7.85	&	1.21	&	154	&-25.5	\\
17-Dec-2003	&	13:21	&	IRTF/MIRSI	&	7.7	&	1.66	&	193	&	-27.0	\\
11-Feb-2007	&	8:50	&	Gemini-S/T-ReCS	&	7.9	&	1.45	&	199	&	-25.4	\\
19-Mar-2007	&	17:08	&	IRTF/MIRSI	&	7.7	&	1.88	&	158	&	-27.5	\\
8-Sep-2007	&	3:40	&	IRTF/MIRSI	&	7.7	&	1.17	&	159	&	26.1	\\
25-Jun-2008	&	11:54	&	Subaru/COMICS	&	7.8	&	1.35	&	161	&	-7.7	\\
15-Sep-2008	&	5:57	&	Subaru/COMICS	&	7.8	&	1.37	&	204	&	24.6	\\
20-Jul-2009	&	1:47	&	IRTF/MIRSI	&	7.7	&	1.21	&	161	&	-12.8	\\
1-Jul-2010	&	12:52	&	IRTF/MIRSI	&	7.7	&	1.52	&	185	&	-26.2	\\
5-Dec-2010	&	5:18	&	IRTF/MIRSI	&	7.7	&	1.09	&	155	&	26.5	\\
16-Jan-2011	&	0:52	&	VLT/VISIR	&	7.9	&	1.55	&	188	&	24.3	\\
16-Jan-2011	&	0:55	&	VLT/VISIR	&	7.9	&	1.55	&	189	&	24.3	\\
30-Oct-2012	&	12:28	&	Subaru/COMICS	&	7.8	&	1.00	&	198	&	-15.8	\\
12-Jan-2017	&	16:13	&	Subaru/COMICS	&	7.8	&	1.13	&	152	&	-27.6	\\
13-Jan-2017	&	12:30	&	Subaru/COMICS	&	7.8	&	2.08	&	167	&	-27.6	\\
13-Jan-2017	&	12:37	&	Subaru/COMICS	&	7.8	&	1.99	&	172	&	-27.6	\\
13-Jan-2017	&	12:42	&	Subaru/COMICS	&	7.8	&	1.93	&	174	&	-27.6	\\
5-Feb-2017	&	15:54	&	Subaru/COMICS	&	7.8	&	1.18	&	154	&	-25.3	\\
18-May-2017	&	9:35	&	Subaru/COMICS	&	7.8	&	1.28	&	170	&	18.0	\\
19-May-2017	&	5:37	&	Subaru/COMICS	&	7.8	&	1.24	&	177	&	18.3	\\
19-May-2017	&	6:10	&	Subaru/COMICS	&	7.8	&	1.16	&	197	&	18.3	\\
27-May-2019	&	10:51	&	Subaru/COMICS	&	7.8	&	1.38	&	157	&	-8.1	\\
3-Jun-2020	&	12:26	&	Subaru/COMICS	&	7.8	&	1.38	&	168	&	-19.0	\\
3-Jun-2020	&	13:06	&	Subaru/COMICS	&	7.8	&	1.33	&	192	&	-19.0	\\
31-Jul-2020 &       10:00       &      Subaru/COMICS	&	7.8	&	1.38	&	179	&	7.8	\\
\lasthline
\end{tabular}
\caption{(continued)}
\end{table*}

\end{document}